\documentstyle[preprint,aps]{revtex}

\tightenlines

\begin{document}

\draft

\title{\bf Series Expansions for the Massive Schwinger Model\\
in Hamiltonian lattice theory}
\author{C.J. Hamer\cite{byline1}, Zheng Weihong\cite{byline2},
 and J. Oitmaa\cite{byline3}} 
\address{School of Physics,                                              
The University of New South Wales,                                   
Sydney, NSW 2052, Australia.}                      

\date{Jan. 16, 1997}

\maketitle 

\begin{abstract}
It is shown that detailed and accurate information about the mass
spectrum of the massive Schwinger model can be obtained using the 
technique of strong-coupling  series expansions. Extended 
strong-coupling series for the energy eigenvalues are calculated,
 and extrapolated to the continuum limit by means of
integrated differential approximants, which are matched onto
a weak-coupling expansion. The numerical estimates are 
compared with exact results, 
and with finite-lattice results calculated for
an equivalent lattice spin model with long-range interactions.
Both the heavy fermion and the light fermion limits
of the model are explored in some detail. 
\end{abstract}                    


\narrowtext
\section{INTRODUCTION}
The Schwinger model\cite{sch,low}, or
quantum electrodynamics in two space-time dimensions,
has been a source of continuing interest over the years. It is a 
fascinating model in its own right, and also exhibits many
of the same phenomena as QCD, such as confinement, chiral
symmetry breaking with a U(1) axial anomaly, and a topological 
$\theta$-vacuum\cite{cas,col76,col75}. It is also perhaps
the simplest non-trivial gauge theory, and this makes it a
standard test-bed for the trial of new techniques for the study
of QCD.

Our main purpose in this paper is to explore the usefulness
of the strong-coupling series approach to this model. It is well
known that
Euclidean Monte Carlo techniques have proved difficult and expensive
to apply to models with dynamical fermions, because of the
infamous ``minus sign'' problem; and thus, 
it seems worthwhile to ask whether other techniques
such as strong-coupling expansions can give useful information in
such cases. Modern linked-cluster techniques\cite{he90}
allow one to carry these expansions to very high order.
We are particularly interested to see if the series
approach can describe the non-relativistic or heavy fermion
limit of these models.

As a first test, we apply the approach to the Schwinger model, 
using a Hamiltonian lattice framework\cite{banks}. We concentrate
on a calculation of the spectrum of bound states, more
specifically the lowest two bound states, for the case of the
massive Schwinger model. It will be shown that the series 
approach can give quite detailed and accurate information on the
spectrum. It will be interesting to see if a similar approach is
useful for models in higher dimensions.

Hamiltonian strong-coupling series were first calculated for the
Schwinger model long ago by Banks, Kogut and Susskind\cite{banks}, 
and were extended by Carroll {\it et al.}\cite{car76}. Since
then, however, the method has fallen into abeyance, being
eclipsed by the power and accuracy of the Monte Carlo
method.

Many other approaches have been made to the bound-state spectrum.
In the massless case, of course, the model is exactly solvable,
as shown by Schwinger\cite{sch,low}. It is equivalent to a
theory of free, massive bosons. For small fermion mass, one
may perturb about the zero-mass limit, and obtain a low-mass
expansion for the spectrum\cite{car76}. This
expansion has recently been carried to second order
by Vary {\it et al.}\cite{var} and Adam\cite{adam}.
In the large mass or non-relativistic limit the ``positronium''
bound states can be solved in terms of a Schr\"{o}dinger
equation with a linear Coulomb potential\cite{ham71}.
For fixed, finite fermion mass, one has to resort to numerical
techniques. A quite accurate variational calculation
in the infinite momentum frame was performed by Bergknoff\cite{ber}.
Finite-lattice Hamiltonian calculation were performed by Crewther and
Hamer\cite{cre80} and Irving and Thomas\cite{irvtho}: 
we include some further finite-lattice
calculations in this paper, mainly as a check on the series results.
Later on Eller, Pauli and Brodsky\cite{ell} applied a ``discrete
light-cone quantization'' (DLCQ) approach to the problem,
and showed that it gave quite good results, not only for
the lowest state, but for a whole range of higher excited states.
Mo and Perry\cite{mo}
have used a different light-front field theory approach, together
with a Tamm-Dancoff  approximation, which appears to give excellent
results. Tomachi and Fujita\cite{tom} have used a ``Bogoliubov
transformation method'', which  works quite well for small
fermion masses.

In Section 2 of the paper the Hamiltonian lattice formulation of 
the model is reviewed, and the known analytic results on
the bound-state spectrum are recalled.
It is shown that for free boundary conditions the gauge
degrees of freedom can be eliminated entirely from the lattice
model, leading to an equivalent spin lattice model with
long-range interactions. This is the formulation which we use
to carry out the finite-lattice calculations.

Section 3 presents the numerical results. Their extrapolation to the
continuum limit is discussed, and expansions near both the low-mass limit
and the non-relativistic limit are analyzed. The numerical results
are compared with previous approaches. Finally our
conclusions are summarized in Section 4.

\section{Formalism}

\subsection{Continuum Formulation}

The continuum Lagrangian density takes the standard form
\begin{equation}
{\cal L} = - {1\over 4} F_{\mu \nu} F^{\mu \nu} + 
\bar{\psi} (i \not\!\partial - g \not\!\!A - m) \psi
\end{equation}
where
\begin{equation}
F_{\mu \nu} = \partial_{\mu} A_{\nu} - \partial_{\nu} A_{\mu}
\end{equation}
and the Lorentz indices $\mu, \nu=0$ or 1. The coupling $g$ in
(1+1) dimensions has the dimensions of mass. Choosing the timelike axial
gauge
\begin{equation}
A_0 =0
\end{equation}
the Hamiltonian is found to be
\begin{equation}
H = \int {\rm d} x \{ - i \bar{\psi} \gamma^1 (\partial_1 + i g A_1 ) \psi + m \bar{\psi} \psi
+ {1\over 2} E^2 \} \label{eq4}
\end{equation}
where the electric field $E$ has only one component in one spatial dimension:
\begin{equation}
E = F^{10}= - \dot{A}^1
\end{equation}

The remaining gauge component  is not an independent degree 
of freedom, but 
can be eliminated if desired, using the constraint provided
by the equation of motion (Gauss' law):
\begin{equation}
\partial_1 E = - \partial_1 \dot{A}^1 = g \bar{\psi} \gamma^0 \psi
\end{equation}

In the massless case, the theory has been solved by Schwinger\cite{sch,low},
and becomes equivalent to a theory of free, massive bosons, with
mass
\begin{equation}
{M_1\over g} = {1\over \sqrt{\pi}}\simeq 0.564
\end{equation}

For small electron mass $m/g$, one can obtain analytic estimates
by perturbing about the massless limit. Caroll, Kogut,
Sinclair and Susskind\cite{car76} found that the lowest-mass (``vector'')
state has mass
\begin{equation}
{M_1\over g} = {1\over \sqrt{\pi}} + e^{\gamma} ({m\over g}) + \cdots
\simeq 0.564 + 1.78 ({m \over g}) +\cdots , \label{eq8}
\end{equation}
while the ratio of the next-lowest (``scalar'') mass to the vector mass  was
\begin{equation}
{M_2\over M_1} = 2 - 2 \pi^3 e^{2\gamma} ({m \over g})^2 + \cdots
\simeq 2 - 197 ({m\over g})^2 +\cdots ,
\end{equation}
where $\gamma\simeq 0.5772...$ is Euler's constant.

These results have been extended to second order by Vary, Fields and
Pirner\cite{var} and by Adam\cite{adam}
\begin{equation}
{M_1\over g} = 0.5642 + 1.781 ({m\over g}) + 0.1907 ({m \over g})^2 
+ \cdots
\end{equation}
Adam\cite{adam} also found
\begin{equation}
{M_2\over M_1} = 2 - {\pi^3 e^{2\gamma} \over 64} ({m\over g})^2 +\cdots
\simeq 2 - 1.5368 ({m\over g})^2 +\cdots
\end{equation}
differing from the result of Carroll {\it et al.}\cite{car76} by
a factor of $2^7=128$.   

In the large-mass or non-relativistic limit,
the ``positronium'' bound states are described by a Schr\"{o}dinger
equation with linear potential\cite{ham71},
\begin{equation}
\left( {p^2\over m} + {1\over 2} g^2 \vert x \vert \right) 
 \Psi (x) = E \Psi (x)\label{eq10}
\end{equation}
where the non-relativistic energy is 
\begin{equation}
E = M - 2 m
\end{equation}
Eq. (\ref{eq10})
may be solved in terms of Airy functions, to
give the energies of the lowest vector and scalar states as\cite{ham71}
\begin{eqnarray}
{E_1\over g} & \sim & 0.642 ({g \over m})^{1/3} \quad {\rm as} \quad m/g \to \infty \nonumber \\
&& \label{eq12} \\
{E_2\over g} & \sim & 1.473 ({g \over m})^{1/3} \quad {\rm as}\quad m/g \to \infty \nonumber 
\end{eqnarray}

\subsection{Lattice Formulation}

The model can now be formulated on a ``staggered'' spatial lattice\cite{banks}.
Let the lattice spacing be $a$, and label the sites with an
integer $n$. Define a single-component fermion field
$\phi (n)$ at each site $n$, obeying anti-commutation relations:
\begin{equation}
\{ \phi^{\dag} (n), \phi (m) \} = \delta_{mn}, \quad 
\{ \phi (n), \phi (m) \} = 0
\end{equation}
The gauge field is defined on the links $(n,n+1)$ connecting each
pair of sites, by
\begin{equation}
U (n ,n+1) = {\rm e}^{i \theta (n) } = {\rm e}^{-iag A^1 (n)}
\end{equation}
Then the lattice Hamiltonian equivalent to Eq.(\ref{eq4}) is
\begin{equation}
H = - {i \over 2 a} \sum_{n=1}^N [ \phi^{\dag} (n) e^{i \theta (n)} 
\phi (n+1) - {\rm h.c.} ] + m \sum_{n=1}^N (-1)^n \phi^{\dag} (n) \phi (n)
+ {g^2 a \over 2} \sum_{n=1}^N L^2 (n) \label{latH}
\end{equation}
where the number of lattice sites $N$ is even, and the correspondence
between lattice and continuum fields is 
\begin{equation}
\phi (n)/\sqrt{a} \to \cases{
\psi_{\rm upper} (x), & $n$ even \cr
\psi_{\rm lower} (x), & $n$ odd \cr }
\end{equation}
(``upper'' and ``lower'' being the two components of the continuum spinor),
and 
\begin{eqnarray}
{1\over ag} \theta (n) &\to& - A^1 (x) \nonumber \\
&& \\
g L(n) &\to& E(x) \nonumber
\end{eqnarray}

The gamma matrices are represented by
\begin{equation}
\gamma^0 = \left( \begin{array}{rr}
1  &0 \\
0  &-1 
\end{array}\right) , \quad
\gamma^1 = \left( \begin{array}{cc}
0  & 1\\
-1 & 0 
\end{array}\right)
\end{equation}
As usual, we have chosen a ``compact'' formulation where the gauge field
becomes an angular variable on the lattice, and $L(n)$ is the 
conjugate `spin' variable
\begin{equation}
[ \theta (n), L (m) ] = i \delta_{nm}
\end{equation}
so that $L(n)$ has integer eigenvalues $L(n)=0,\pm 1,\pm 2, \cdots$.
As noted by Banks {\it et al.}\cite{banks}, this quantization of
electric field (or flux) in one dimension also occurs in 
the continuum Schwinger model, due to Gauss' law. If one takes the
naive continuum limit $a\to 0$, the lattice Hamiltonian (\ref{latH})
reduces to the continuum Hamiltonian (\ref{eq4}), as it should.

Now define the dimensionless operator
\begin{equation}
W = {2\over ag^2} H = W_0 + x V
\end{equation}
where
\begin{eqnarray}
W_0 &=& \sum_n L^2 (n) + \mu \sum_n (-1)^n \phi^{\dag} (n) \phi (n) \\
V &=& -i \sum_n [ \phi^{\dag} (n) e^{i \theta (n)} \phi (n+1) - {\rm h.c.}]\\
\mu &=& {2m\over g^2 a}, \quad x = {1\over g^2 a^2}
\end{eqnarray}

For $x\ll 1$ one can employ strong-coupling perturbation theory on this
model, treating $W_0$ as the unperturbed Hamiltonian and $V$ as
the perturbation, as discussed by Banks {\it et al.}\cite{banks}.
In the strong-coupling limit, the unperturbed ground 
state $\vert 0\rangle$ is the
eigenstate with
\begin{equation}
L(n) =0, \quad \phi^{\dag} (n) \phi (n) = {1\over 2} [ 1 - (-1)^n ], \quad {\rm all~~} n
\end{equation}
whose energy will be normalized to zero, corresponding to 
a ``filled Dirac sea''.
Banks {\it et al.}\cite{banks} have discussed how to use 
Rayleigh-Schr\"{o}dinger
perturbation theory to generate perturbation series in $x$ for
the ground state and excited state eigenvalues of this system, and the 
discussion will not be repeated here. We have used more sophisticated
linked-cluster techniques\cite{he90} to generate high-order perturbation series
for these eigenvalues, as presented below.

The lattice version of Gauss' law is then taken as
\begin{equation}
L(n) - L(n-1) = \phi^{\dag} (n) \phi (n) - {1\over 2} [ 1 - (-1)^n ]
\end{equation}
which means excitations on odd and even sites create $\mp 1$ units of 
flux, corresponding
to `electron' and `positron' excitations, respectively.

\subsection{Equivalent spin formulation}

The one-component fermion operators can be replaced by 
Pauli spin operators at each site if we employ a Jordan-Wigner 
transformation\cite{banks},
\begin{eqnarray}
\phi (n) &=& \prod_{l<n} [ i \sigma_3 (l) ] \sigma^- (n) \\
\phi^{\dag} (n) &=& \prod_{l<n} [- i \sigma_3 (l) ] \sigma^{+} (n) 
\end{eqnarray}
giving 
\begin{eqnarray}
W_0 &=& \sum_n L^2 (n) + {\mu \over 2} \sum_n (-1)^n \sigma_3 (n) + N\mu/2 \\
V &=& \sum_n [ \sigma^{+} (n) e^{i \theta (n)} \sigma^- (n+1) + {\rm h.c.} ]
\end{eqnarray}
The strong-coupling ground-state then corresponds to
\begin{equation}
L(n) =0, \quad \sigma_3 (n) = - (-1)^n, \quad {\rm all~} n
\end{equation}
Next, the gauge field can be eliminated using Gauss' law:
\begin{equation}
L(n) - L(n-1) = {1\over 2} [\sigma_3 (n) + (-1)^n ]
\end{equation}
and a residual gauge transformation:
\begin{equation}
\sigma^- (n) \to \prod_{l<n} \{ {\rm e}^{-i \theta (l)} \} \sigma^- (n)
\end{equation}
provided that we assume free boundaries
\begin{equation}
L(0)=L(N)=0, \quad {\rm where~} N= \# {\rm ~lattice~sites}
\end{equation}
[If periodic boundary conditions are assumed, then there is
one extra  independent gauge degree of freedom left over, corresponding
to the ``background'' electric field\cite{col76}].
The resulting Hamiltonian is then
\begin{equation}
W = W_0 + x V \label{eqH}
\end{equation}
where
\begin{eqnarray}
W_0 &=& {\mu \over 2} \sum_{n=1}^N (-1)^n \sigma_3 (n) + {N \mu\over 2}
+ \sum_{n=1}^{N-1} \left[ {1\over 2} \sum_{m=1}^n ( \sigma_3 (m) + (-1)^m )
\right] ^2 \label{eq34} \\
V &=& \sum_{n=1}^{N-1} [ \sigma^{+} (n) \sigma^- (n+1) + {\rm h.c.} ] \label{eq35}
\end{eqnarray}
All trace of the gauge field has now disappeared, 
but instead there is a non-local,
long-range interaction between the spins in 
the last term of equation (\ref{eq34}),
which of course corresponds to the long-range Coulomb interaction
between charges in the original theory.
In the continuum limit $a\to 0$, $x\to \infty$, the interaction $V$
dominates the Hamiltonian so that in leading order the system becomes
equivalent to a simple XY model with ground-state energy per site
\begin{equation}
{\omega_0\over N} \to  - {x\over \pi} \quad {\rm as} \quad x\to \infty
\label{eq36}
\end{equation}
The energy gap to the lowest excited state in the sector of
``vector'' states will correspond to the vector mass, so we expect
\begin{equation}
\omega_1 - \omega_0 =
{2\over a g^2} M_1 = 
{2\over g}M_1\sqrt{x}  \label{eq37}
\end{equation}
and similarly in the scalar or ground-state sector, the minimum energy gap
\begin{equation}
\omega_2 -\omega_0 =
{2\over a g^2} M_2 = 
{2\over g}M_2\sqrt{x} \label{eq38}
\end{equation}
Our aim in this paper is to find estimates of these masses $M_1$ 
and $M_2$.

The equivalent spin model has a total of only $2^N$ 
possible configurations,
and lends itself readily to finite-lattice techniques of 
analysis. As a check on the series results, we have used the Lanczos
algorithm to obtain exact results for the low-lying eigenvalues
of this system at finite $x$ on lattices of up to $N=22$ sites.
The ground-state energy $\omega_0$ is easily obtained as the lowest
eigenvalue in the sector containing the unperturbed ground-state
$\vert 0\rangle$; and the first excited state energy $\omega_1$ is 
likewise the lowest eigenvalue in the ``vector'' state sector,
corresponding in the strong-coupling limit to the state\cite{banks}
\begin{equation}
\vert 1 \rangle = {1\over \sqrt{N}} \sum_{n-1}^{N-1} 
\left[ \sigma^{+} (n) \sigma^- (n+1) - {\rm h.c.} \right] \vert 0 \rangle
\end{equation}
The second excited-state energy $\omega_2$ is 
the lowest of a ``band'' of excited states
in the vacuum sector, corresponding in the strong-coupling limit to
the state 
\begin{equation}
\vert 2 \rangle = {1\over \sqrt{N}} \sum_{n=1}^{N-1}
\left[ \sigma^{+} (n) \sigma^{-} (n+1) + {\rm h.c.} \right] \vert 0 \rangle ~.
\label{s_sta}
\end{equation}

\section{Results and Analysis}

\subsection{Finite-lattice results}
Exact eigenvalues have been calculated for the equivalent spin
Hamiltonian (\ref{eqH})  using the Lanczos technique
for various values of $m/g$ and  coupling $x$, on even lattices from
$N=4$ up to $N=22$ sites. No symmetrization of states was
employed: since free boundary conditions were chosen, the
system does not exhibit translational invariance in any
case. The calculations are carried out in the sector 
$\sum_i \sigma_3 (i)=0$ which has $N!/[(N/2)!]^2$ states, 
and the ground-state energy $\omega_0$ and the
``vector" excited-state energy $\omega_1$ are the lowest and the second lowest
eigenvalues in this sector, respectively. 
The ``scalar" excited-state energy $\omega_2$ is
slightly more tricky to obtain  because there are several other states,
corresponding to the momentum excitations of the ``vector" excited state,
in this sector giving lower energy than the ``scalar" excited state.
We have got over this problem
by moving in stages from the strong-coupling limit 
$x=0$ to the desired coupling value, and ``tagging'' the
desired state as that which has maximum overlap with its 
predecessor at each stage, starting from the state $\vert 2 \rangle$ in (\ref{s_sta}).

To make comparison with series data, it is first necessary to 
extrapolate the finite-lattice data to the bulk limit
$N\to \infty$. The convergence of the finite-lattice sequences
was slow. Tests showed that the convergence was polynomial
in $1/N$: thus for the ground-state energy
\begin{equation}
\omega_0 (N) \to
N \epsilon_0 (\infty ) + b_0 + {b_1\over N^2} + O(N^{-3})\quad {\rm as} \quad N\to \infty
\end{equation}
where $\epsilon_0 (\infty )$ is the bulk energy per site and 
$b_0$ is the surface energy term;
while for the energy gaps
\begin{equation}
\omega_i (N) - \omega_0 (N)  \to
b_0' + { b_1'\over N^2} + O(N^{-3}) \quad {\rm as} \quad N\to \infty
  \label{mfin}
\end{equation}
This is opposed to the exponential convergence
$\sim \exp (-c N)$ one normally expects for periodic
boundary conditions. There is a simple explanation
for this behaviour\footnote{We are indebted to Dr. O. Sushkov for this
remark.}: on the lattice
with free boundaries, the excitations have finite momentum
$O(\pi/N)$, and thus their energies include a kinetic energy correction
term $O((\pi/N)^2)$. 
As one approaches $x\to \infty$, which is a critical point of the
lattice model, the finite-lattice corrections become relatively 
much larger, and extrapolation to the bulk limit becomes progressively
more difficult.
An example is shown in Figure 1 for $x=4, m/g=0$.

Various sequence extrapolation algorithms\cite{bar,hen,gut}
were tried for estimating the bulk limit, such as the
alternating VBS algorithm, and the Lubkin and Bulirsch-Stoer algorithms,
but the most reliable and accurate method in this case seemed
to be a simple least squares polynomial fit in $1/N$,
with fictitious errors assigned to the data points. 
Consistency between the estimates at different orders allows a
crude estimate of the likely error in the final result.
This was the technique used to obtain the estimates
used in the rest of this paper.

\subsection{Series expansions}
Strong-coupling perturbation series have been calculated for the ground-state
energy $\omega_0$, and the energy gaps ($\omega_1-\omega_0$) and
($\omega_2-\omega_0$), as functions of the coupling $x$ and mass 
parameter $\mu$.
The full series up to order $x^{10}$ are presented in the Appendix.

These series can be analyzed in three different regimes, as follows:

\subsubsection{Massless limit, $m/g\to 0$}

For small $m/g$, the eigenvalues can be expanded as series in the
mass parameter $\mu$, e.g. for the vector gap
\begin{equation}
(\omega_1 - \omega_0 ) [x,\mu] - 2 \mu = 
f_0 (x) + \mu f_1 (x) + \mu^2 f_2 (x) + \cdots
\end{equation}
The series have been calculated for $f_i(x)$ ($i=0,1,\cdots,6$) up to order
$x^{28}$ for the vector excited state and order $x^{26}$ for the scalar excited state.
Coefficients for the series $f_i (x)$ ($i=0,1,2,3$) are listed in Table I.
The expected
behaviour in the continuum
limit for these series is
\begin{equation}
f_l (x) \to a_l x^{(1-l)/2} \quad {\rm as} \quad x\to \infty \label{eq40}
\end{equation}
which would lead to a continuum energy gap
\begin{equation}
{M - 2 m \over g} = a_0/2 + a_1 (m/g) + 2 a_2 (m/g)^2 + \cdots \label{eq41}
\end{equation}
These series must now be extrapolated from the strong-coupling
limit $x=0$ to the continuum limit $x\to\infty$.
We have employed the standard techniques of integrated
differential approximants and naive Pad\'{e} 
approximants\cite{gut} for this purpose, combined with
a ``matching'' technique. Some examples
are shown in Figs. 2, 3 and 4.

Figure 2 shows the ground-state energy per site for $m/g=0$ as
a function of $y=1/\sqrt{x}$ (which is the natural variable
to use at weak couplings, as shown by equation (\ref{eq40})
and by previous weak-coupling analyses\cite{ham71,ken}). 
It can be seen that the  approximants
converge down to about $y\simeq 0.5$, and are in excellent 
 agreement with the 
finite-lattice estimates. These can easily be extrapolated to
the continuum limit $x=0$, where 
\begin{equation}
{\omega_0\over Nx} = - {1\over \pi}
\end{equation}
according to equation (\ref{eq36}).

A more interesting example is seen in Figure 3, which shows the
vector energy gap as a function of $y$. There it can be
seen that the series approximants converge down to about
$y\simeq 0.7$, and then begin to spray outwards; however,
the established behaviour is quite smooth as a function of
$y$, and a polynomial fit in powers of $y$ to
the series data over the range 0.7-1.1 gives a ``matching''
weak-coupling series fit 
which extrapolates to a value

\begin{equation}
{M_1 - 2 m \over g}
\to  0.56(2) \quad {\rm as} \quad y\to 0
\end{equation}
which compares well with the expected exact value of 0.5642, from
equations (\ref{eq8}) and (\ref{eq37}). It can also be seen that the
finite-lattice estimates converge well down to $y\simeq 0.2$, confirming
that the extrapolation to weak coupling is valid.  A fit to the finite-lattice
data gives an even more accurate estimate of the continuum
limit
\begin{equation}
{M_1 - 2 m \over g}\to  0.57(1)\quad {\rm as} \quad y\to 0
\end{equation}

Figure 4 is a similar plot for the energy of the scalar excited state.
Here it can be seen that the series approximants converge down to
$y\simeq 1$, where the function develops a pronounced peak or bump.
It is not really possible to tell from the diverging series
approximants whether the function increases, decreases or remains
flat below that point, but the finite-lattice data show that it 
decreases gently towards the exact continuum value 1.128.
A fit to the finite-lattice data over 0.2-0.8 in $y$ gives 
\begin{equation}
{M_2 - 2 m \over g} \to  1.14(3)\quad {\rm as} \quad y\to 0
\end{equation}
in quite reasonable agreement with the exact value; but
the best one could do with the series data would be to 
estimate a qualitative value,
\begin{equation}
a_0^{\rm scalar} \simeq 1.25(15).
\end{equation}

In a similar fashion, estimates have been obtained for 
the coefficients
$a_0$, $a_1$, $a_2$, $a_3$, in equation (\ref{eq41}) for the vector and
scalar masses, extracted from both the series data and the 
finite-lattice data. These are summarized in Table II, along with
the exact results (where known). It can be seen that the
numerical estimates are in good agreement with the
exact results for the vector state, and so the estimate for
$a_3$ should also be fairly reliable. For the scalar state,
there is more structure at small $y$, as seen in Figure 4,  and
although the estimates for $a_0$ and $a_1$ agree quite well with
the exact results, those for $a_2$ and $a_3$ are virtually worthless.

\subsubsection{Finite $m/g$}

At fixed, finite $m/g$, 
the series have been calculated 
up to order $x^{30}$ for the ground-state energy $\omega_0$,
and order $x^{53/2}$ for the energy gaps
 $(\omega_1 -\omega_0)$ and $(\omega_2 -\omega_0)$ -
these series are available on request.
The analysis follows very similar
lines to those described above. The major difference is that
for $m/g\not= 0$, the strong-coupling series expansions are
in powers of $x^{1/2}$, rather than $x^2$. The series are
therefore longer, but the convergence is not very different to that at $m/g=0$.

At large $m/g$ the series coefficients begin to
grow very rapidly, like $({m/g})^{n}$ at large orders
$n$, and it becomes more difficult to maintain good
accuracy down to small values of $y$.
Figures 5 and 6 illustrate the behaviour of
the energy gap at large $m/g$.
It can be seen that any structure has moved to smaller values of $y$,
and is somewhat less pronounced than at $m/g=0$.

Our estimates of the energies $E_i/g$ are shown in Table III and
Figure 7, along with earlier finite-lattice estimates of 
Crewther and Hamer\cite{cre80}, and the light-cone estimates of 
Eller {\it et al.}\cite{ell} and Mo and Perry\cite{mo}. It can
be seen that all the estimates agree with each other, within
errors, except that the results of Eller {\it et al.}\cite{ell}
run a little too high at small $m/g$. For the vector state, the
data match on  beautifully to the asymptotic expansions
at both ends. For the scalar state, the data indicate
a peak in the energy at about $m/g\simeq 0.5$.

Comparing the different estimates, it is noticeable that
our current finite-lattice estimates are in fact 
less  accurate than the old estimates of 
Crewther and Hamer\cite{cre80} or the even more accurate 
results of Irving and Thomas\cite{irvtho},
although there is good consistency
between them. This is because of the large finite-size
corrections associated with the free boundary conditions.
More importantly, the series estimates are seen to be in good
agreement with the current finite-lattice estimates, and
for the vector state they are generally almost as accurate
(approximately 10\%), although for the scalar state they are
worse (approximately 15-20\%). The results of Mo and Perry\cite{mo}
seem very reliable and accurate, but unfortunately there is no
assessment of the likely errors in their results.

\subsubsection{Non-relativistic limit, $m/g\to \infty$}

At large $m/g$, a natural re-arrangement of the 
series for the energy gaps was suggested by Hamer\cite{ham71}.
Instead of variables $\mu$ and $x$, one can re-arrange the series in terms
of variables $1/\mu$ and $u$, where
\begin{equation}
u = {x^2\over \mu} = {1\over 2 m g^2 a^3}={1\over 2 m/g} x^{3/2}
\end{equation}
Then one finds the series can easily be expanded in powers of the
inverse mass parameter $1/\mu$, e.g. for the vector gap
\begin{equation}
(\omega_1 - \omega_0) [u,\mu] - 2 \mu =
\tilde{f}_0 (u) + {1\over \mu} \tilde{f}_1 (u)
+ {1\over \mu^2} \tilde{f}_2 (u) +\cdots \label{seru}
\end{equation}
The continuum limit can then be approached by first letting
$m/g$ (or $\mu$) $\to\infty$ in such a way that 
$mg^2$ (or $u$) remains finite, and then letting $a\to 0$, or
$u\to \infty$. If we {\it assume} that each separate term
on the right-hand side of (\ref{seru}) gives a finite contribution
to the continuum energy gap in this limit, then we require an asymptotic
behaviour for the $\tilde{f}_l (u)$ given by
\begin{equation}
\tilde{f}_l (u) \to
c_l u^{(l+1)/3} \quad {\rm as} \quad u\to \infty\label{asymu}
\end{equation}
which would lead to a continuum energy gap
\begin{equation}
{E \over g} = \sum_{l=0}^\infty {c_l\over 2} 
\left( {g \over 2 m}\right)^{(4l+1)/3} \label{expu}
\end{equation}
The leading term $O(({g/m})^{1/3})$ has the correct behaviour
as predicted by (\ref{eq12}).

The series have been calculated for $\tilde{f}_l$ ($l=0,1,\cdots,6$) up to order
$u^{14}$ for the vector excited state and order $u^{13}$ for the scalar excited state.
Coefficients for the series $\tilde{f}_l (u)$ ($l=0,1,2,3$)  are listed
in Table IV for the two energy gaps. The first five coefficients
of $\tilde{f}_0 (u)$ can be obtained from the results of
Carroll {\it et al.}\cite{car76} and were listed previously by
Hamer\cite{ham71}. The series for $\tilde{f}_0 (u)$ is identical
with that obtained from a lattice version of the non-relativistic
Schr\"{o}dinger equation (\ref{eq10}), as shown by
Kenway and Hamer\cite{ken}, which confirms that we should obtain
at least the leading term in the non-relativistic limit correctly
by this procedure.

The series for $\tilde{f}_l (u)$ must now be extrapolated
from the strong-coupling limit $u=0$ to the continuum limit
$u\to \infty$. The same techniques were employed for this
purpose as for the previous series in $x$. The only question is
what to take as a weak-coupling variable. No detailed weak-coupling
analysis has been done in the
non-relativistic limit; but empirical tests on $\tilde{f}_0 (u)$,
making comparison with the exact results, appear to show that
the best extrapolations are obtained using $z=1/u$ as weak-coupling
variable. The results are shown in Figures 8 and 9.

For the vector state, it can be seen that the integrated differential
approximants converge down to $z\simeq 0.2$, and the function behaves
very smoothly in $z$, so that a very accurate extrapolation to
the continuum limit is possible, giving
\begin{equation}
c_0^{\rm vector} = 1.618(2)
\end{equation}
which compares well with the expected exact  value of
$1.617(=2^{4/3}\times 0.642)$. 
Note that the series results allow a more accurate estimate for
this quantity than the finite-lattice data do.

The results for the scalar state are shown in Figure 9, where it can
be seen that the series approximants converge only down to
$z\simeq 0.7$, but the function is again quite smooth in $z$, so
that a fairly reliable extrapolation may be made to the continuum limit,
as confirmed by the finite lattice data. The resulting estimate
is 
\begin{equation}
c_0^{\rm scalar}=3.73(3)
\end{equation}
to be compared with the exact value of $3.71(=2^{4/3}\times 1.473$).

As soon as one tries to go beyond the leading order, however,
some major problems arise. An example is shown in Figure 10, which
shows $\tilde{f}_1 (u)/u^{2/3}$ for the vector state. It can be
seen that the series approximants and the finite-lattice estimates agree
very well down to very small values of $z$, and indicate a divergent
behaviour for this quantity. A Dlog Pad\'{e} analysis of the series indicates
that $\tilde{f}_1 (u)$ behaves asymptotically more like $u$ than $u^{2/3}$.
Similar results are obtained for all the $\{ \tilde{f}_l (u), l>0\}$, 
for both the vector and scalar states.

It therefore appears that the assumption
(\ref{asymu}) is incorrect, and that the terms in equation (\ref{seru})
cannot be analyzed separately (beyond the leading order, at least).
It is very likely that the limiting behaviour of the energy gaps is
non-uniform, and that by letting first $m/g\to \infty$,
and then $a\to 0$, we have taken the limits in the wrong order.
This gives an incorrect result for the ground-state energy, for instance.
It follows that the expansion (\ref{expu}) is also incorrect; and
in fact one would probably expect higher-order  corrections to involve
integer powers of $(g/m)$, rather than $(g/m)^{4/3}$. It is an interesting puzzle
how to obtain useful estimates of the higher-order corrections from
the series data in these circumstances.
At this stage, we  have no answer to the puzzle.

\section{Conclusions}

It has been shown that strong-coupling series expansions can
deliver quite detailed and accurate information about the mass spectrum
of the Schwinger model. Using integrated differential approximants
or other methods, the series can be continued or extrapolated
well into the weak-coupling regime, and then matched onto a weak-coupling
form, which allows reasonably accurate estimates of the continuum
limit. This has been confirmed by comparison with both finite-lattice
data and exact results.
At fixed, finite fermion mass the estimates for the
lowest, vector excited state energy were accurate
to about
10 percent. For the next lowest, scalar excited state, the 
eigenvalue shows more structure at weak coupling, particularly
for small
$m/g$, and the continuum estimates are more qualitative, at
about the 15-20 percent level.

Estimates can also be obtained for the expansion coefficients of the
continuum energy eigenvalues in powers of $(m/g)$ about the zero-mass
limit, $m/g\to 0$. These numerical estimates agree with the exact
results of Carroll {\it et al.}\cite{car76},
Vary {\it et al.}\cite{var} and Adam\cite{adam}, and extend them
by one order. The series results for these coefficients have 
an accuracy equal to or better than that of
the finite-lattice results.

A complementary expansion was attempted about the non-relativistic
limit, $m/g\to \infty$.
Series estimates for the leading-order term of the non-relativistic energy
were extremely accurate in this limit, with an accuracy of 0.2\% for
the vector state, but problems arose for the higher-order
corrections.
There appears to be a non-uniform
limiting behaviour in this case, and the natural structure of the
strong-coupling series does not predict the correct limiting behaviour.
We have not resolved the puzzle of how to analyze the series in
this situation. It would be very interesting to compare the 
series data with  more detailed analytic calculations in the
non-relativistic limit - we hope to address this problem in
the future.

Finite-lattice techniques can also give  accurate information
about this model, as previously demonstrated by 
Crewther and Hamer\cite{cre80} and Irving and Thomas\cite{irvtho}.
Exact finite-lattice eigenvalues have been calculated for the
equivalent spin Hamiltonian on lattices of up to $N=22$ sites, and
then extrapolated to the bulk limit by polynomial fits in 
$1/N$. Fitting these estimates to a weak-coupling form, estimates 
of the continuum limit can be obtained which are accurate to about
5-10 percent for both the vector and scalar states.
This is actually  worse than the accuracy of the old
finite-lattice data of Crewther and Hamer\cite{cre80}.

The factor which hindered the finite-lattice calculations
from giving even more accurate estimates was the large size
of the finite-lattice corrections,
$O(1/N^2)$, which were associated with the free boundary conditions.
In retrospect, it might be better to choose periodic boundary
conditions. This carries the penalty that the gauge field cannot
be entirely eliminated, and one is left with one extra gauge
degree of freedom corresponding to Coleman's background
electric field or flux, which would have to be truncated in some 
fashion. The advantage would be that the finite-lattice corrections
should be much smaller, and the finite-lattice sequence should 
converge more rapidly. With today's computers, one could
probably expect to obtain virtually exact eigenvalues on
lattices up to some 20 sites, and thus obtain a substantial
improvement on previous finite-lattice calculations\cite{cre80,irvtho}.

In higher dimensions, the finite-lattice approach is hardly
feasible, because of the huge proliferation of
basis states with lattice size. Monte Carlo techniques are the only
other option in this direction, and their power is still
rather limited for models with dynamical fermions.
It remains to be seen whether modern series expansion
techniques can deliver useful information for these more complicated models.

\acknowledgments
This work forms part of a research project supported by a grant 
from the Australian Research Council. 

\appendix
\section*{}

The full strong coupling series up to order $x^{10}$ are
\begin{eqnarray}
\omega_0/N &=&  -{{{x^2}}\over {1 + 2\,\mu }} + 
   {{3\,{x^4}}\over {{{\left( 1 + 2\,\mu  \right) }^3}}} - 
   {{2\,\left( 29 + 20\,\mu  \right) \,{x^6}}\over 
     {{{\left( 1 + 2\,\mu  \right) }^5}\,\left( 3 + 2\,\mu  \right) }} 
 + {{\left( 1443 + 2000\,\mu  + 700\,{{\mu }^2} \right) \,{x^8}}\over 
     {{{\left( 1 + 2\,\mu  \right) }^7}\,{{\left( 3 + 2\,\mu  \right) }^2}}} \nonumber \\
&& - {{4\,\left( 51093 + 126936\,\mu  + 117230\,{{\mu }^2} + 
         47484\,{{\mu }^3} + 7056\,{{\mu }^4} \right) \,{x^{10}}}\over 
     {{{\left( 1 + 2\,\mu  \right) }^9}\,{{\left( 3 + 2\,\mu  \right) }^3}\,
       \left( 5 + 2\,\mu  \right) }}
\end{eqnarray}

\begin{eqnarray}
\omega_1 - \omega_0 &&=
  1 + 2\,\mu  + {{2\,{x^2}}\over {1 + 2\,\mu }} - 
   {{2\,\left( 5 + 2\,\mu  \right) \,{x^4}}\over 
     {{{\left( 1 + 2\,\mu  \right) }^3}}} + 
   {{4\,\left( 59 + 68\,\mu  + 24\,{{\mu }^2} + 4\,{{\mu }^3} \right) \,
       {x^6}}\over 
     {{{\left( 1 + 2\,\mu  \right) }^5}\,\left( 3 + 2\,\mu  \right) }} 
\nonumber \\
 && - 
   {{2\,\left( 3313 + 6056\,\mu  + 3934\,{{\mu }^2} + 1188\,{{\mu }^3} + 
         216\,{{\mu }^4} + 16\,{{\mu }^5} \right) \,{x^8}}\over 
     {{{\left( 1 + 2\,\mu  \right) }^7}\,{{\left( 3 + 2\,\mu  \right) }^2}}} \nonumber \\
&& +  2 ( 
3578209 + 11359410 \mu  + 14982934 {{\mu }^2} + 
         10681694 {{\mu }^3} + 4498120 {{\mu }^4} + 1134808 {{\mu }^5}  \nonumber \\
 &&     +   148448 {{\mu }^6} - 1888 {{\mu }^7} - 3712 {{\mu }^8} - 
         384 {{\mu }^9}
  )  {x^{10}} [
     {
{{\left( 1 + 2 \mu  \right) }^9} {{\left( 3 + 2 \mu  \right) }^3} 
       \left( 5 + 2 \mu  \right)  \left( 7 + 2 \mu  \right) 
}]^{-1}
\end{eqnarray}

\begin{eqnarray}
\omega_2 - \omega_0 &&=
  1 + 2\,\mu  + {{6\,{x^2}}\over {1 + 2\,\mu }} - 
   {{2\,\left( 13 + 2\,\mu  \right) \,{x^4}}\over 
     {{{\left( 1 + 2\,\mu  \right) }^3}}} + 
   {{4\,\left( 143 + 112\,\mu  + 4\,{{\mu }^2} - 4\,{{\mu }^3} \right) \,
       {x^6}}\over 
     {{{\left( 1 + 2\,\mu  \right) }^5}\,\left( 3 + 2\,\mu  \right) }} 
\nonumber \\
 && + 
   {{2\,\left( -7905 - 11632\,\mu  - 4510\,{{\mu }^2} + 28\,{{\mu }^3} + 
         104\,{{\mu }^4} - 16\,{{\mu }^5} \right) \,{x^8}}\over 
     {{{\left( 1 + 2\,\mu  \right) }^7}\,{{\left( 3 + 2\,\mu  \right) }^2}}} \nonumber \\
&& + ( 17055678 + 48682964 \mu  + 54776904 {{\mu }^2} + 
         30230052 {{\mu }^3} + 8173504 {{\mu }^4} + 987088 {{\mu }^5} \nonumber \\
&&    + 
         157952 {{\mu }^6} + 74944 {{\mu }^7} + 14336 {{\mu }^8} + 
         768 {{\mu }^9} )  {x^{10}}  [
     {{{\left( 1 + 2 \mu  \right) }^9} {{\left( 3 + 2 \mu  \right) }^3} 
       \left( 5 + 2 \mu  \right)  \left( 7 + 2 \mu  \right) }]^{-1}
\end{eqnarray}


\begin{figure}
\caption{Finite-size scaling behaviour of the vector energy gap
$(M_1-2m)/g$ as a function of $1/N^2$ for $m/g =0$, $x=4$, where $N$ is
the lattice size. Solid boxes mark the finite-lattice data points; 
the dashed 
line shows a polynomial fit of the form (\protect\ref{mfin}).}
\label{fig:fig1}
\end{figure}

\begin{figure}
\caption{\protect Ground-state energy per site 
${\omega_0/(xN)}$ as a function of $y=1/\protect\sqrt{x}$ for $m/g=0$,
The dashed lines are integrated differential approximants to the series
data, while the solid boxes with error bars represent finite-lattice estimates.
The solid line is a fit to the finite-lattice data in powers of $y$.
The exact continuum limit is marked by an open circle.}
\label{fig:fig2}
\end{figure}

\begin{figure}
\caption{Vector mass gap $M_1/g$ as a function of 
$y=1/\protect\sqrt{x}$ for $m/g=0$. Dashed lines are series approximants,
solid boxes are finite-lattice data, and the open 
circle marks the exact continuum limit. The solid line is a 
linear fit in powers of $y$ to the finite-lattice data.}
\label{fig:fig3}
\end{figure}

\begin{figure}
\caption{Scalar mass gap $M_2/g$
as a function of  $y=1/\protect\sqrt{x}$,
for $m/g=0$. Notation as Fig. 3. Here the solid line represents
a second-order polynomial fit to the finite-lattice data over the range $[0.2,0.8]$ in $y$.}
\label{fig:fig4}
\end{figure}

\begin{figure}
\caption{The vector energy gap 
${(M_1-2m)/g}$
as a function of $y=1/\protect\sqrt{x}$ for $m/g=32$.
Notation as Fig. 4. 
Here the solid line represents a second-order polynomial fit to the finite-lattice
data over the range [0-0.5] in $y$.}
\label{fig:fig5}
\end{figure}

\begin{figure}
\caption{The scalar energy gap ${(M_2-2m)/g}$ as a function of 
$y=1/\protect\sqrt{x}$ for $m/g=32$. Notation as Fig. 4.
Here the solid line represents a  polynomial fit to the finite-lattice
data over the range [0-0.7] in $y$.}
\label{fig:fig6}
\end{figure}

\begin{figure}
\caption{Estimates of the bound-state energies ${E_i/g}$ versus ${m/g}$
for both the vector and scalar states. The open circles are our series
estimates, the solid boxes are our finite-lattice estimates, the
dashed line is the result of Mo and Perry, and the solid lines represent the asymptotic
behaviour in the limiting case $m/g\to 0$, and the non-relativistic limit.}
\label{fig:fig7}
\end{figure}

\begin{figure}
\caption{Graph of the quantity ${\tilde{f}_0/u^{1/3}}$ as a function of
 $z=1/u$ for the vector excited state. Notation as in Fig. 4.
The solid line represents a polynomial fit in $1/u$ to the series results.}
\label{fig:fig8}
\end{figure}

\begin{figure}
\caption{Graph of the quantity ${\tilde{f}_0/u^{1/3}}$ as a function of
 $z=1/u$ for the scalar
excited state. Notation as Fig. 4. The solid line is a linear  fit in
$z=1/u$ to the series data over the range [0.6-1] in $z$.}
\label{fig:fig9}
\end{figure}

\begin{figure}
\caption{Graph of the quantity ${\tilde{f}_1/u^{2/3}}$ as a function of
 $z=1/u$ for the vector
excited state. The solid lines are integrated differential approximants to the
series data, and the solid boxes are finite-lattice estimates.}
\label{fig:fig10}
\end{figure}

\setdec 0.000000000000000
\begin{table}
\squeezetable
\caption{Series coefficients in $x$ for  the series 
$f_0$, $f_1$, $f_2$, and $f_3$ for the
vector and scalar excited states. Nonzero coefficients of $x^n$ are listed.
}\label{tab1}
\begin{tabular}{rrrrr}
\multicolumn{1}{c}{$n$} &\multicolumn{1}{c}{$f_0$}
&\multicolumn{1}{c}{$f_1$} &\multicolumn{1}{c}{$f_2$}&\multicolumn{1}{c}{$f_3$} \\
\hline
\multicolumn{5}{c}{vector excited state}\\
  0 &\dec  1.000000000000 &\dec  0.000000000000 &\dec  0.000000000000 &\dec  0.000000000000 \\
  2 &\dec  2.000000000000 &\dec $-$4.000000000000 &\dec  8.000000000000 &\dec $-$1.600000000000$\times 10^{1}$ \\
  4 &\dec $-$1.000000000000$\times 10^{1}$ &\dec  5.600000000000$\times 10^{1}$ &\dec $-$2.160000000000$\times 10^{2}$ &\dec  7.040000000000$\times 10^{2}$ \\
  6 &\dec  7.866666666667$\times 10^{1}$ &\dec $-$7.484444444444$\times 10^{2}$ &\dec  4.344296296296$\times 10^{3}$ &\dec $-$1.979753086420$\times 10^{4}$ \\
  8 &\dec $-$7.362222222222$\times 10^{2}$ &\dec  9.942962962963$\times 10^{3}$ &\dec $-$7.742029629630$\times 10^{4}$ &\dec  4.547973004115$\times 10^{5}$ \\
 10 &\dec  7.572929100529$\times 10^{3}$ &\dec $-$1.326103683144$\times 10^{5}$ &\dec  1.296902835617$\times 10^{6}$ &\dec $-$9.337217941993$\times 10^{6}$ \\
 12 &\dec $-$8.273669056689$\times 10^{4}$ &\dec  1.780034473768$\times 10^{6}$ &\dec $-$2.096087983379$\times 10^{7}$ &\dec  1.787437096185$\times 10^{8}$ \\
 14 &\dec  9.428034196036$\times 10^{5}$ &\dec $-$2.405247329390$\times 10^{7}$ &\dec  3.312148486580$\times 10^{8}$ &\dec $-$3.264352340840$\times 10^{9}$ \\
 16 &\dec $-$1.108357531764$\times 10^{7}$ &\dec  3.269962317907$\times 10^{8}$ &\dec $-$5.154702578031$\times 10^{9}$ &\dec  5.764697147243$\times 10^{10}$ \\
 18 &\dec  1.334636403738$\times 10^{8}$ &\dec $-$4.469545711783$\times 10^{9}$ &\dec  7.935843765541$\times 10^{10}$ &\dec $-$9.927975671713$\times 10^{11}$ \\
 20 &\dec $-$1.637995781331$\times 10^{9}$ &\dec  6.137741867815$\times 10^{10}$ &\dec $-$1.211925759935$\times 10^{12}$ &\dec  1.676878382095$\times 10^{13}$ \\
 22 &\dec  2.041592824001$\times 10^{10}$ &\dec $-$8.462460202654$\times 10^{11}$ &\dec  1.839263212769$\times 10^{13}$ &\dec $-$2.788695534036$\times 10^{14}$ \\
 24 &\dec $-$2.577314968112$\times 10^{11}$ &\dec  1.170804957950$\times 10^{13}$ &\dec $-$2.777427143130$\times 10^{14}$ &\dec  4.579160745014$\times 10^{15}$ \\
 26 &\dec  3.288641969992$\times 10^{12}$ &\dec $-$1.624677659161$\times 10^{14}$ &\dec  4.176990920608$\times 10^{15}$ &\dec $-$7.439926786605$\times 10^{16}$ \\
 28 &\dec $-$4.234675197596$\times 10^{13}$ &\dec  2.260335118290$\times 10^{15}$ &\dec $-$6.260315585321$\times 10^{16}$ &\dec  1.197974308422$\times 10^{18}$ \\
\hline
\multicolumn{5}{c}{scalar excited state}\\
  0 &\dec  1.000000000000 &\dec  0.000000000000 &\dec  0.000000000000 &\dec  0.000000000000 \\
  2 &\dec  6.000000000000 &\dec $-$1.200000000000$\times 10^{1}$ &\dec  2.400000000000$\times 10^{1}$ &\dec $-$4.800000000000$\times 10^{1}$ \\
  4 &\dec $-$2.600000000000$\times 10^{1}$ &\dec  1.520000000000$\times 10^{2}$ &\dec $-$6.000000000000$\times 10^{2}$ &\dec  1.984000000000$\times 10^{3}$ \\
  6 &\dec  1.906666666667$\times 10^{2}$ &\dec $-$1.884444444444$\times 10^{3}$ &\dec  1.120829629630$\times 10^{4}$ &\dec $-$5.195753086420$\times 10^{4}$ \\
  8 &\dec $-$1.756666666667$\times 10^{3}$ &\dec  2.435066666667$\times 10^{4}$ &\dec $-$1.932472592593$\times 10^{5}$ &\dec  1.151850271605$\times 10^{6}$ \\
 10 &\dec  1.804833650794$\times 10^{4}$ &\dec $-$3.218263683144$\times 10^{5}$ &\dec  3.192825551667$\times 10^{6}$ &\dec $-$2.325598645777$\times 10^{7}$ \\
 12 &\dec $-$1.979052000756$\times 10^{5}$ &\dec  4.314452729791$\times 10^{6}$ &\dec $-$5.136573365113$\times 10^{7}$ &\dec  4.420974571815$\times 10^{8}$ \\
 14 &\dec  2.267367521095$\times 10^{6}$ &\dec $-$5.842838007004$\times 10^{7}$ &\dec  8.115577786905$\times 10^{8}$ &\dec $-$8.058540815871$\times 10^{9}$ \\
 16 &\dec $-$2.681007957489$\times 10^{7}$ &\dec  7.972653477157$\times 10^{8}$ &\dec $-$1.265571251416$\times 10^{10}$ &\dec  1.424081244509$\times 10^{11}$ \\
 18 &\dec  3.246557142476$\times 10^{8}$ &\dec $-$1.094270467476$\times 10^{10}$ &\dec  1.954139217902$\times 10^{11}$ &\dec $-$2.457363530118$\times 10^{12}$ \\
 20 &\dec $-$4.005343740541$\times 10^{9}$ &\dec  1.508952103990$\times 10^{11}$ &\dec $-$2.994043095570$\times 10^{12}$ &\dec  4.161067716407$\times 10^{13}$ \\
 22 &\dec  5.016016326942$\times 10^{10}$ &\dec $-$2.088733926173$\times 10^{12}$ &\dec  4.558824467106$\times 10^{13}$ &\dec $-$6.938732610121$\times 10^{14}$ \\
 24 &\dec $-$6.359434181405$\times 10^{11}$ &\dec  2.900477258173$\times 10^{13}$ &\dec $-$6.905937254404$\times 10^{14}$ &\dec  1.142452124172$\times 10^{16}$ \\
 26 &\dec  8.145971243132$\times 10^{12}$ &\dec $-$4.038508440296$\times 10^{14}$ &\dec  1.041673344896$\times 10^{16}$ &\dec $-$1.861014699378$\times 10^{17}$ \\
\end{tabular}
\end{table}

\setdec 0.00000000000
\begin{table}
\squeezetable
\caption{Table of estimates of $a_0,~ a_1,~ \cdots,~a_3$
for the vector and scalar masses, obtained from series data,
finite-lattice data, and exact results\protect\cite{var,adam}.}\label{tab2}
\begin{tabular}{rrrrr}
\multicolumn{1}{c}{} & \multicolumn{1}{c}{$a_0/2$} 
 & \multicolumn{1}{c}{$a_1+2$} & \multicolumn{1}{c}{$2a_2$} 
 & \multicolumn{1}{c}{$4a_3$} \\
\hline
\multicolumn{5}{c}{Vector state} \\
Series &  0.56(2)  & 1.80(2)   & 0.16(4)  & -0.22(6)  \\
Finite lattice &  0.57(1)  & 1.78(2)   & 0.3(1)   &   \\
Exact  & 0.5642 & 1.781 & 0.1908 &   \\
\multicolumn{5}{c}{Scalar state} \\
Series &  1.25(15)  & 3.2(2)   & -4(2)   &   \\
Finite lattice&  1.14(3) & 2.5(10)  & 0(2)   &   \\
Exact  & 1.1284 & 3.562 & -0.485 &   \\
\end{tabular}
\end{table}

\setdec 0.00000000000
\begin{table}
\squeezetable
\caption{Estimates of the continuum bound-state energies
$E_1/g$, $E_2/g$ as functions of $m/g$. 
The series and finite-lattice estimates obtained in this
work are compared with the earlier finite-lattice estimates
of Crewther and Hamer\protect\cite{cre80}, and the light-cone estimates
of Eller {\it et al.}\protect\cite{ell} and Mo and Perry\protect\cite{mo}.
}\label{tab3}
\begin{tabular}{rrrrrr}
\multicolumn{1}{c}{} & \multicolumn{1}{c}{series} 
 & \multicolumn{1}{c}{finite-lattice} & \multicolumn{1}{c}{C \& H} 
 & \multicolumn{1}{c}{Eller {\it et al.}} & \multicolumn{1}{c}{Mo \& Perry} \\
\multicolumn{1}{c}{$m/g$} & \multicolumn{1}{c}{this work} & \multicolumn{1}{c}{this work}
& \multicolumn{1}{c}{\cite{cre80}} & \multicolumn{1}{c}{\cite{ell}}& \multicolumn{1}{c}{\cite{mo}} \\
\hline
\multicolumn{6}{c}{Vector state} \\
0     &  0.56(2)    &  0.57(1) &  0.56(1)   &         &         \\
0.125 &  0.53(3)    &  0.52(2) &  0.54(1)   &    0.60 &    0.54 \\
0.25  &  0.52(4)    &  0.52(2) &  0.52)1)   &    0.53 &    0.52 \\
0.5   &  0.50(2)    &  0.50(2) &  0.50(1)   &    0.49 &    0.49 \\
1     &  0.46(4)    &  0.46(3) &  0.46(1)   &    0.44 &    0.44\\
2     &  0.41(4)    &  0.41(3) &  0.413(5)  &    0.39 &    0.39\\
4     &  0.34(2)    &  0.35(2) &  0.358(5)  &    0.34 &    0.34\\
8     &  0.30(3)    &  0.31(2) &  0.299(5)  &    0.28 &    0.29\\
16    &  0.24(3)    &  0.25(2) &  0.245(5)  &    0.23 &    0.24\\
32    &  0.20(4)    &  0.20(2) &  0.197(5)  &    0.20 &    0.20\\
\hline
\multicolumn{6}{c}{Scalar state} \\
0     &  1.25(15) &   1.14(3) &  1.12(5) &             &      \\
0.125 &  1.35(15) &   1.24(4) &  1.11(5) &      1.41   &  1.22\\
0.25  &  1.30(15) &   1.27(5) &  1.12(5) &      1.31   &  1.23\\
0.5   &  1.25(15) &   1.25(5) &  1.15(5) &      1.23   &  1.20\\
1     &  1.10(15) &   1.14(5) &  1.19(5) &      1.13   &  1.12\\
2     &  1.00(15) &   1.01(5) &  1.10(5) &      0.98   &  0.99\\
4     &  0.90(15) &   0.85(4) &  0.93(5) &      0.84   &  0.84\\
8     &  0.70(15) &   0.73(5) &  0.77(5) &      0.69   &  0.70\\
16    &  0.5(1)   &   0.59(5) &  0.62(5) &      0.55   &  0.56\\
32    &  0.4(1)   &   0.50(5) &  0.49(5) &      0.46   &  0.46\\
\end{tabular}
\end{table}

\setdec 0.0000000000000000
\begin{table}
\squeezetable
\caption{Series coefficients in $u$ for  the series $\tilde{f}_0$, $\tilde{f}_1$,  $\tilde{f}_2$, and  $\tilde{f}_3$ 
for the vector and scalar excited states.
}\label{tab4}
\begin{tabular}{rrrrr}
\multicolumn{1}{c}{$n$} &\multicolumn{1}{c}{$\tilde{f}_0$}
&\multicolumn{1}{c}{$\tilde{f}_1$} &\multicolumn{1}{c}{$\tilde{f}_2$}
&\multicolumn{1}{c}{$\tilde{f}_3$} \\
\hline
\multicolumn{5}{c}{vector excited state}\\
  0 &\dec  1.000000000000 &\dec  0.000000000000 &\dec  0.000000000000 &\dec  0.000000000000 \\
  1 &\dec  1.000000000000 &\dec $-$5.000000000000$\times 10^{-1}$ &\dec  2.500000000000$\times 10^{-1}$ &\dec $-$1.250000000000$\times 10^{-1}$ \\
  2 &\dec $-$5.000000000000$\times 10^{-1}$ &\dec $-$5.000000000000$\times 10^{-1}$ &\dec  1.125000000000 &\dec $-$1.250000000000 \\
  3 &\dec  2.500000000000$\times 10^{-1}$ &\dec  5.000000000000$\times 10^{-1}$ &\dec  6.875000000000$\times 10^{-1}$ &\dec $-$3.437500000000 \\
  4 &\dec $-$6.250000000000$\times 10^{-2}$ &\dec $-$4.375000000000$\times 10^{-1}$ &\dec $-$6.718750000000$\times 10^{-1}$ &\dec $-$1.375000000000 \\
  5 &\dec $-$4.687500000000$\times 10^{-2}$ &\dec  2.500000000000$\times 10^{-1}$ &\dec  6.484375000000$\times 10^{-1}$ &\dec  1.496093750000 \\
  6 &\dec  7.161458333333$\times 10^{-2}$ &\dec $-$3.906250000000$\times 10^{-3}$ &\dec $-$5.179036458333$\times 10^{-1}$ &\dec $-$1.298828125000 \\
  7 &\dec $-$3.765190972222$\times 10^{-2}$ &\dec $-$1.729600694444$\times 10^{-1}$ &\dec  2.191840277778$\times 10^{-1}$ &\dec  9.108886718750$\times 10^{-1}$ \\
  8 &\dec $-$1.051613136574$\times 10^{-2}$ &\dec  1.898419415509$\times 10^{-1}$ &\dec  1.635357892072$\times 10^{-1}$ &\dec $-$4.226029007523$\times 10^{-1}$ \\
  9 &\dec  3.595046055170$\times 10^{-2}$ &\dec $-$6.325276692708$\times 10^{-2}$ &\dec $-$4.165481755763$\times 10^{-1}$ &\dec $-$1.153888466917$\times 10^{-1}$ \\
 10 &\dec $-$2.772084302863$\times 10^{-2}$ &\dec $-$9.454201278373$\times 10^{-2}$ &\dec  3.571410159515$\times 10^{-1}$ &\dec  5.813870214140$\times 10^{-1}$ \\
 11 &\dec  6.427749717514$\times 10^{-4}$ &\dec  1.602973539774$\times 10^{-1}$ &\dec $-$1.180466277773$\times 10^{-2}$ &\dec $-$7.371812834432$\times 10^{-1}$ \\
 12 &\dec  2.087420768938$\times 10^{-2}$ &\dec $-$9.161565485249$\times 10^{-2}$ &\dec $-$3.598007444620$\times 10^{-1}$ &\dec  3.895490578301$\times 10^{-1}$ \\
 13 &\dec $-$2.186375166505$\times 10^{-2}$ &\dec $-$4.596852371627$\times 10^{-2}$ &\dec  4.496844531498$\times 10^{-1}$ &\dec  3.420860280769$\times 10^{-1}$ \\
 14 &\dec  5.564121895836$\times 10^{-3}$ &\dec  1.366704793607$\times 10^{-1}$ &\dec $-$1.603366455727$\times 10^{-1}$ &\dec $-$9.530272768917$\times 10^{-1}$ \\
\hline
\multicolumn{5}{c}{scalar excited state}\\
  0 &\dec  1.000000000000 &\dec  0.000000000000 &\dec  0.000000000000 &\dec  0.000000000000 \\
  1 &\dec  3.000000000000 &\dec $-$1.500000000000 &\dec  7.500000000000$\times 10^{-1}$ &\dec $-$3.750000000000$\times 10^{-1}$ \\
  2 &\dec $-$5.000000000000$\times 10^{-1}$ &\dec $-$2.500000000000 &\dec  4.125000000000 &\dec $-$4.250000000000 \\
  3 &\dec $-$2.500000000000$\times 10^{-1}$ &\dec  1.250000000000 &\dec  3.562500000000 &\dec $-$1.187500000000$\times 10^{1}$ \\
  4 &\dec $-$6.250000000000$\times 10^{-2}$ &\dec  8.125000000000$\times 10^{-1}$ &\dec $-$4.046875000000 &\dec $-$4.187500000000 \\
  5 &\dec  4.687500000000$\times 10^{-2}$ &\dec  1.718750000000$\times 10^{-1}$ &\dec $-$2.632812500000 &\dec  1.437109375000$\times 10^{1}$ \\
  6 &\dec  7.161458333333$\times 10^{-2}$ &\dec $-$3.554687500000$\times 10^{-1}$ &\dec $-$9.993489583333$\times 10^{-2}$ &\dec  8.261718750000 \\
  7 &\dec  3.765190972222$\times 10^{-2}$ &\dec $-$4.944661458333$\times 10^{-1}$ &\dec  2.158637152778 &\dec $-$2.571695963542 \\
  8 &\dec $-$1.051613136574$\times 10^{-2}$ &\dec $-$2.319607204861$\times 10^{-1}$ &\dec  2.582842791522 &\dec $-$1.177163357205$\times 10^{1}$ \\
  9 &\dec $-$3.595046055170$\times 10^{-2}$ &\dec  1.738077799479$\times 10^{-1}$ &\dec  8.375302538460$\times 10^{-1}$ &\dec $-$1.176848630552$\times 10^{1}$ \\
 10 &\dec $-$2.772084302863$\times 10^{-2}$ &\dec  3.925574031877$\times 10^{-1}$ &\dec $-$1.691848582201 &\dec $-$9.678567015095$\times 10^{-1}$ \\
 11 &\dec $-$6.427749717514$\times 10^{-4}$ &\dec  2.711228939731$\times 10^{-1}$ &\dec $-$2.874944640017 &\dec  1.281942563814$\times 10^{1}$ \\
 12 &\dec  2.087420768938$\times 10^{-2}$ &\dec $-$6.325043120993$\times 10^{-2}$ &\dec $-$1.598536929541 &\dec  1.730423290807$\times 10^{1}$ \\
 13 &\dec  2.186375166505$\times 10^{-2}$ &\dec $-$3.223758205015$\times 10^{-1}$ &\dec  1.178489577243 &\dec  6.362423273963 \\
\end{tabular}
\end{table}

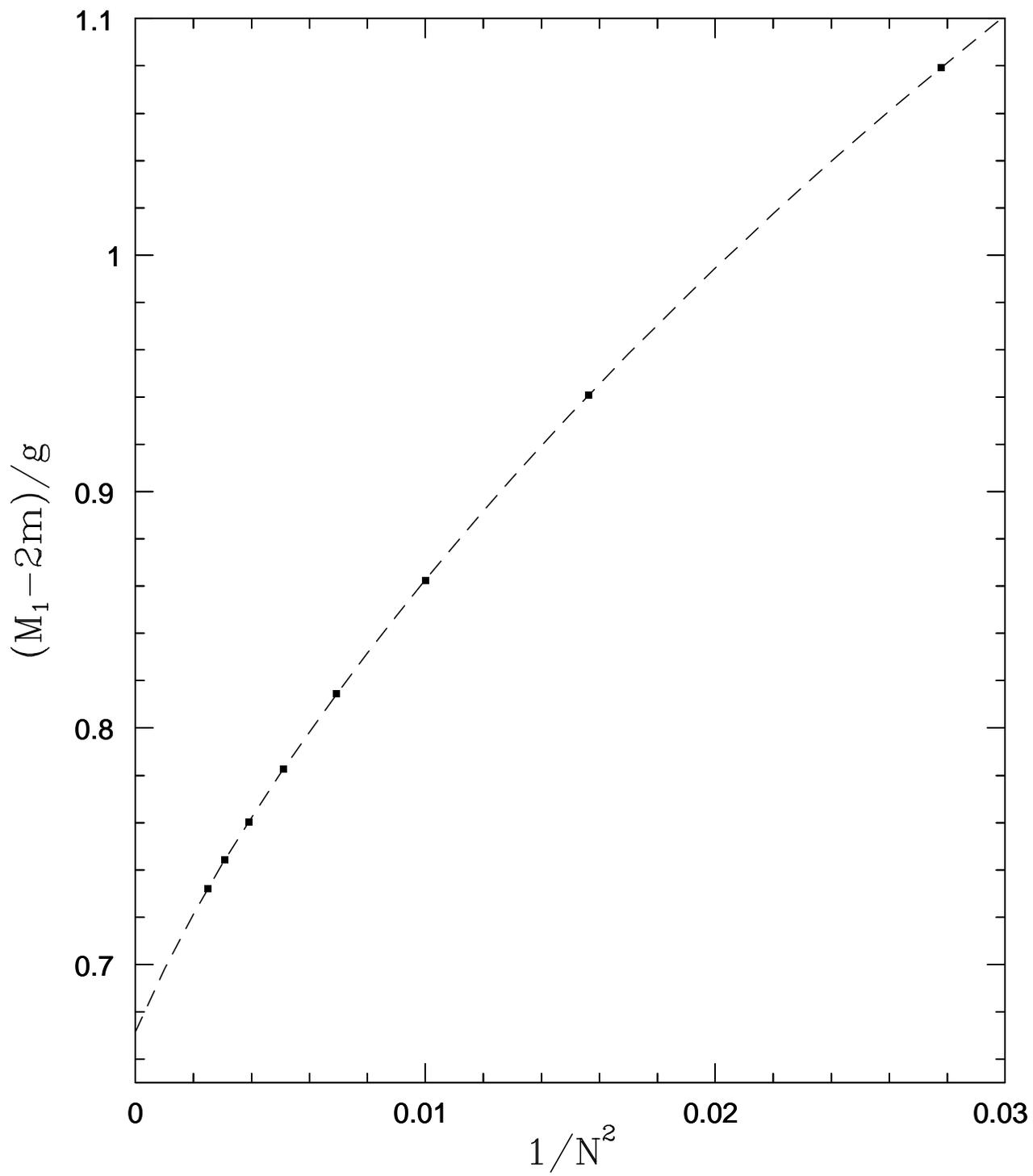
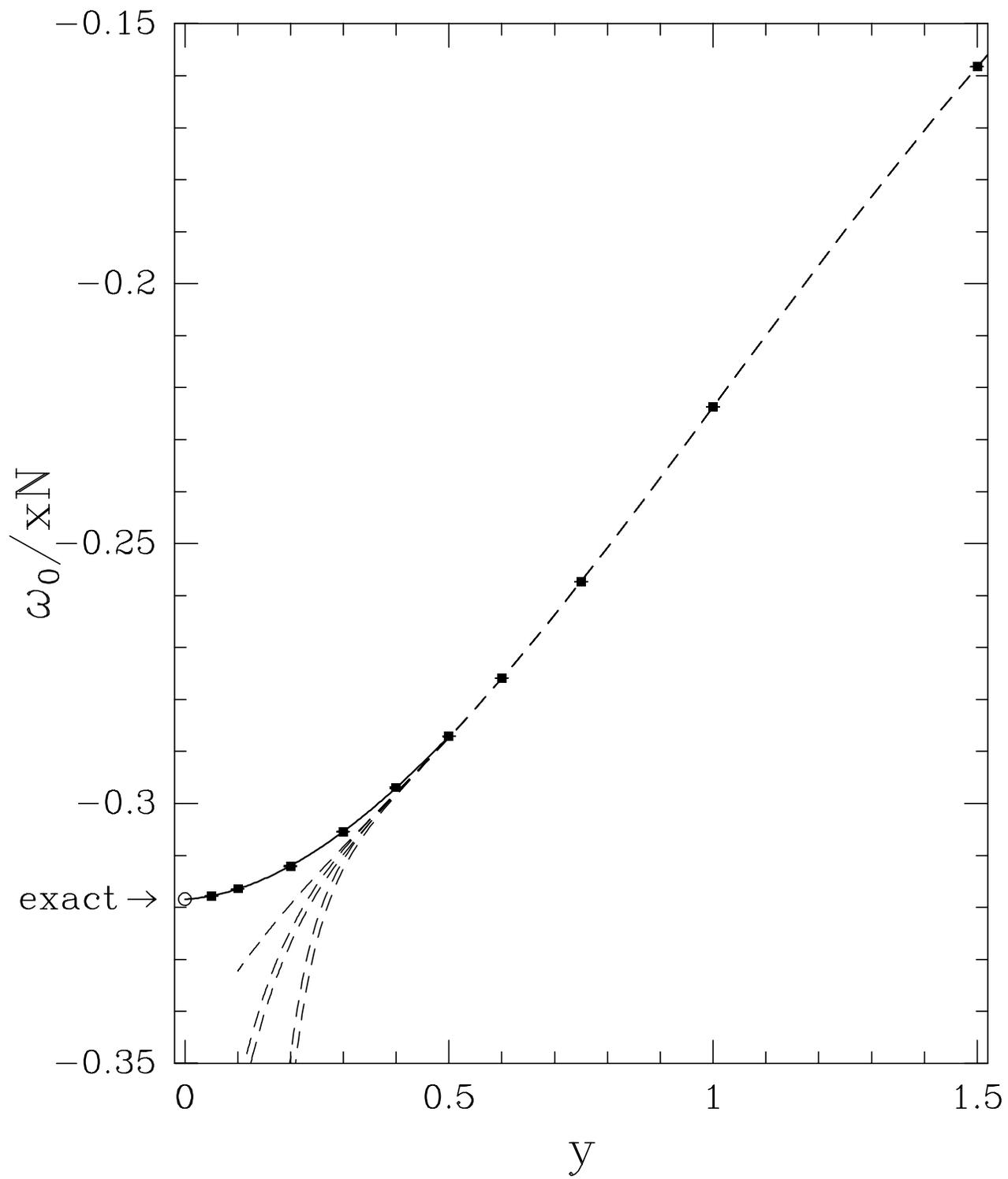
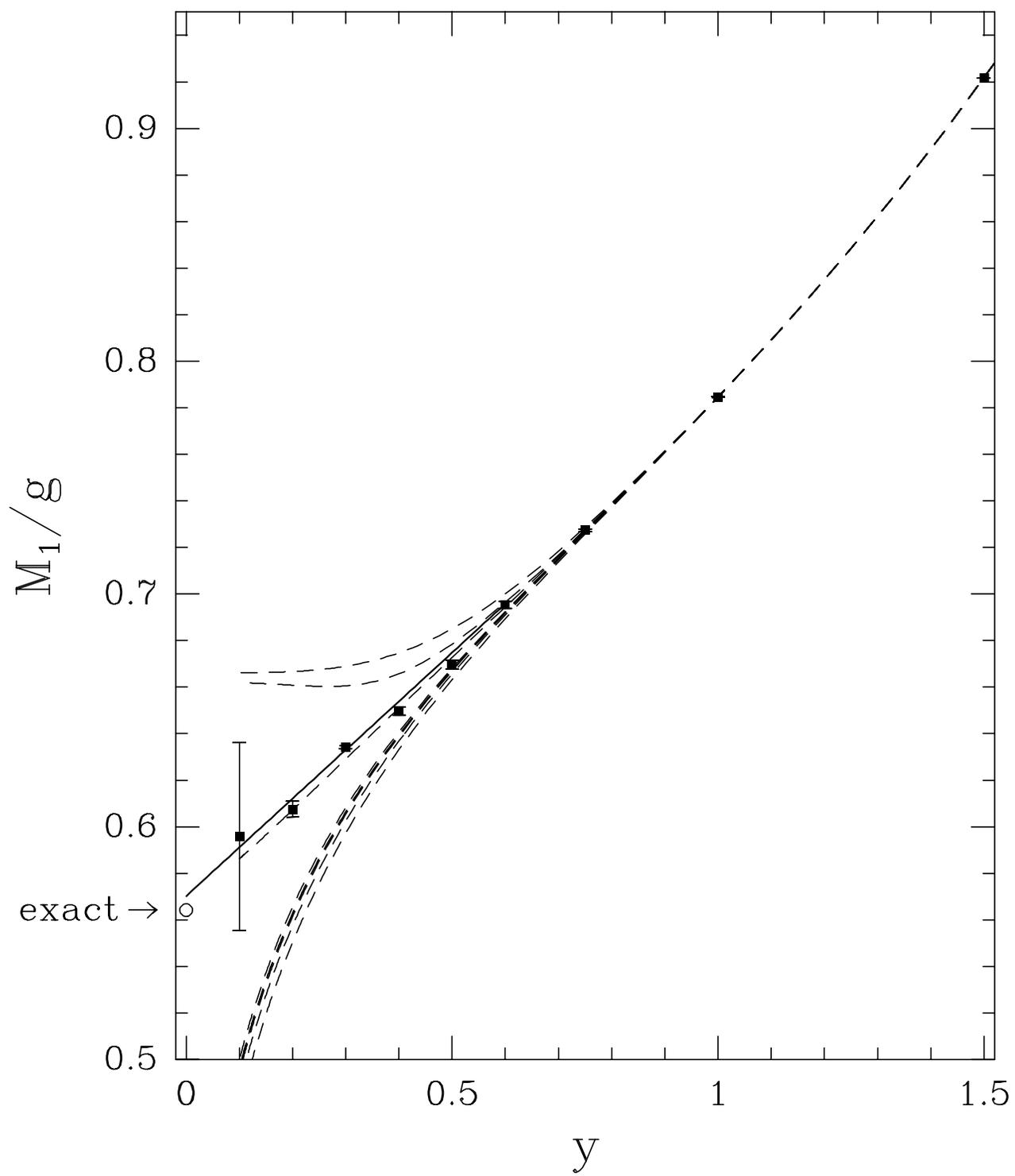
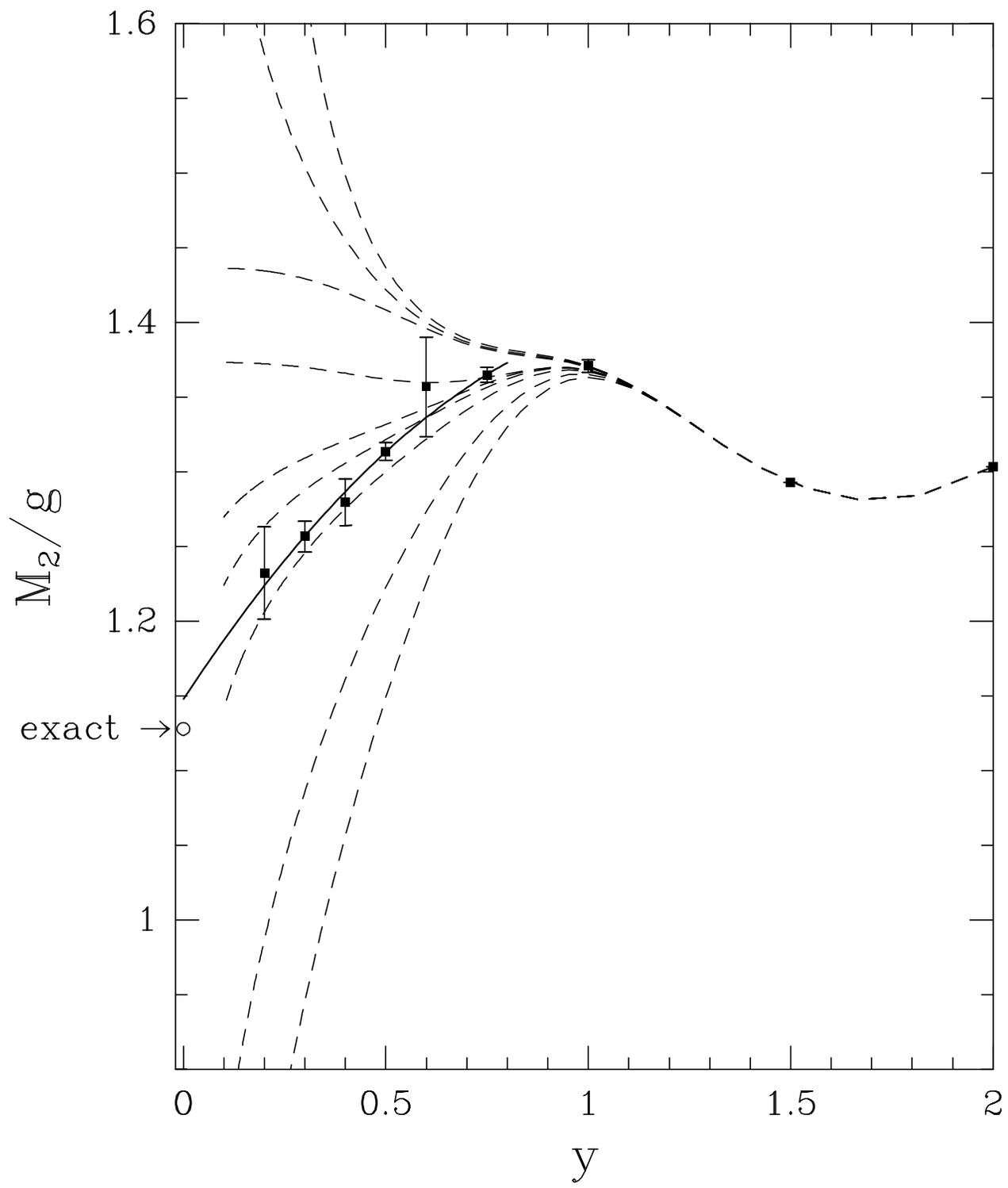
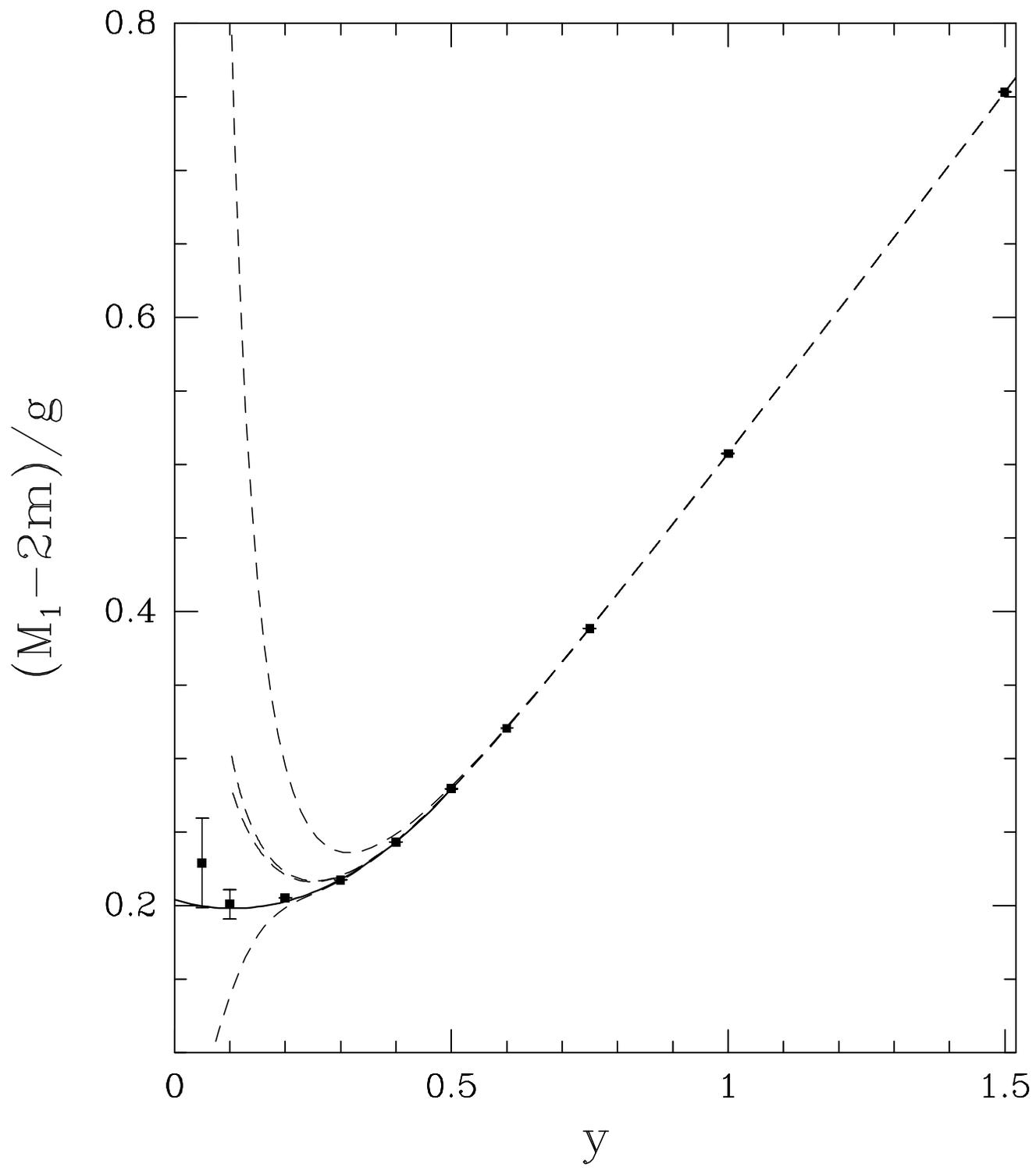
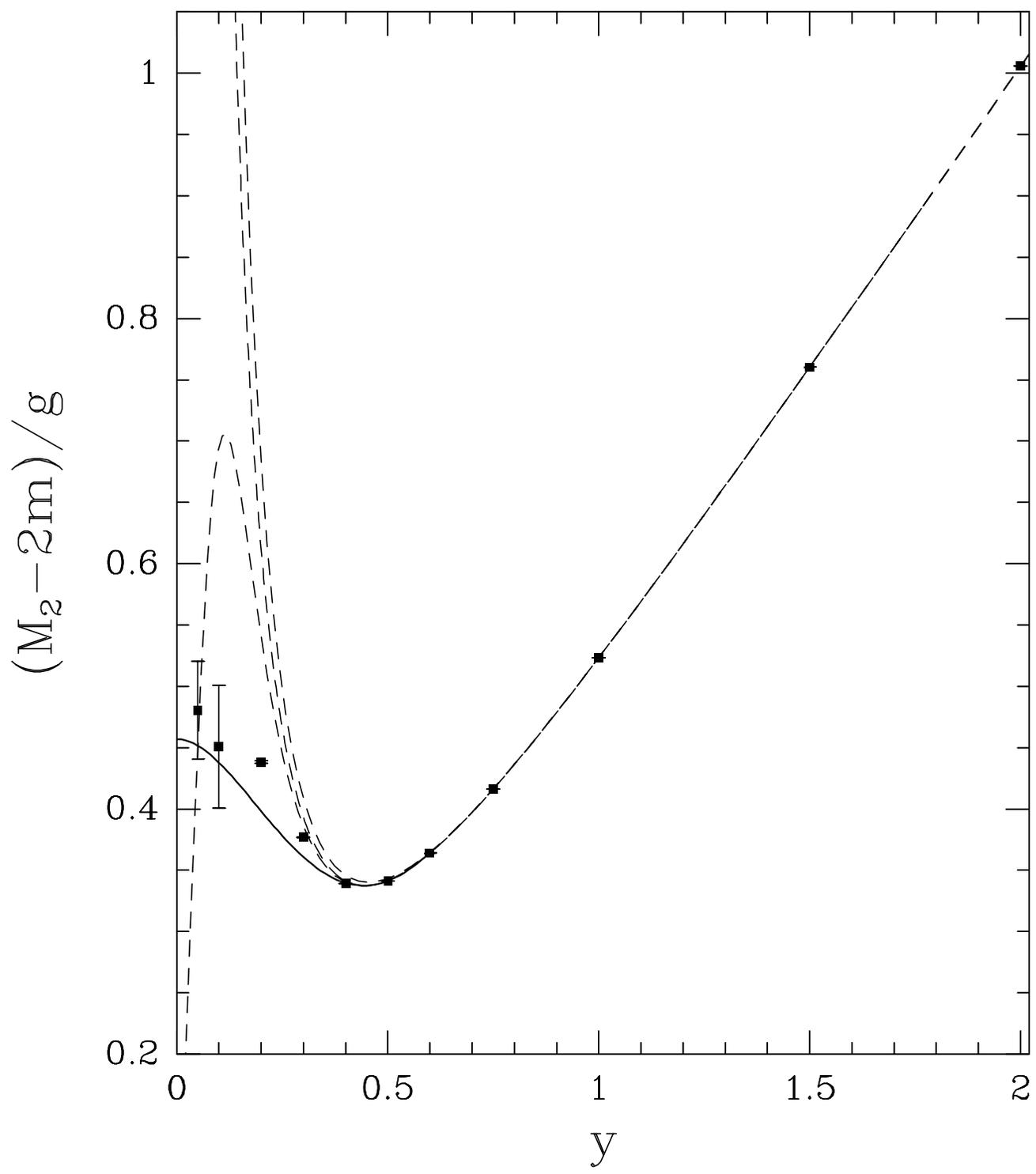
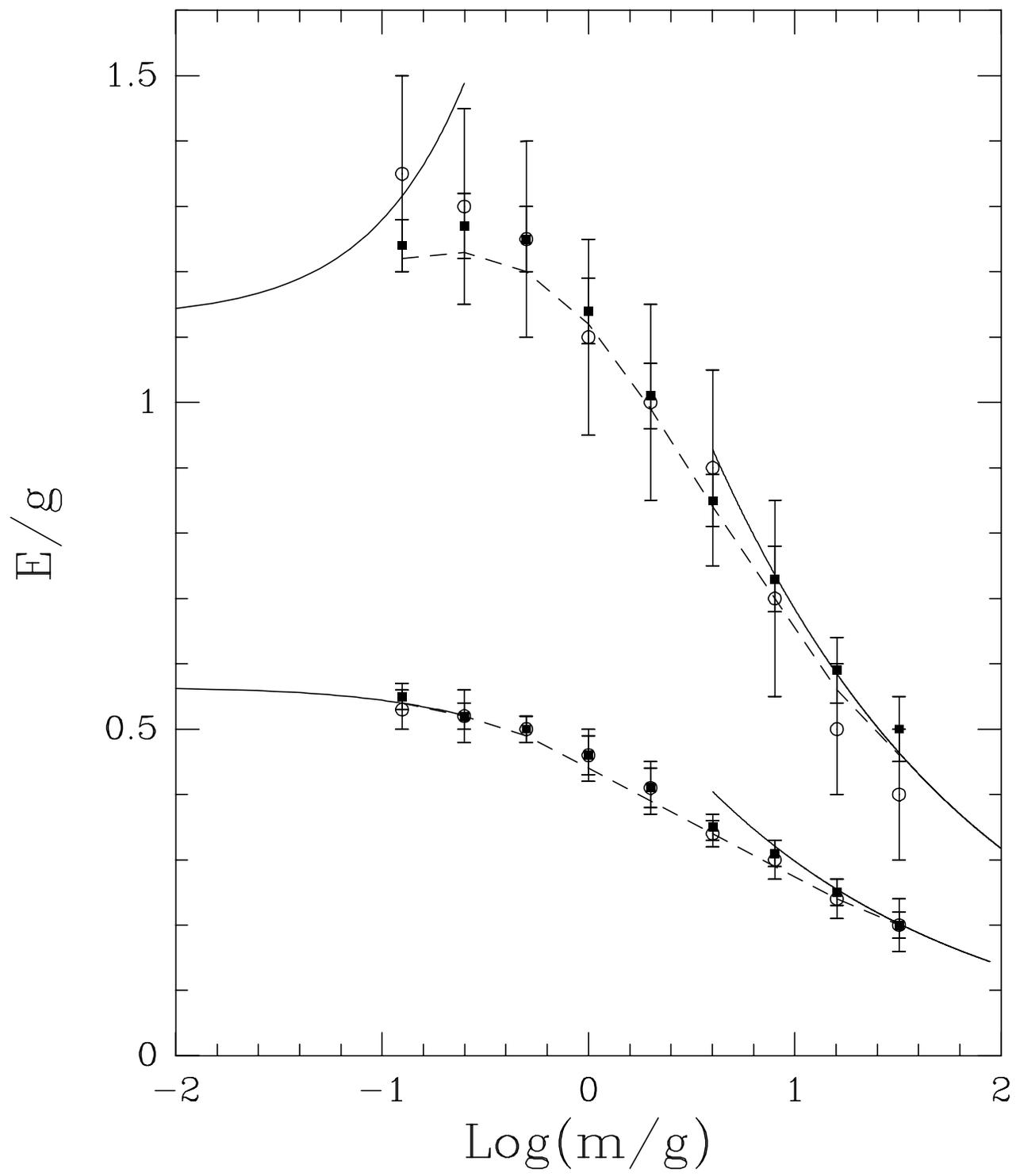
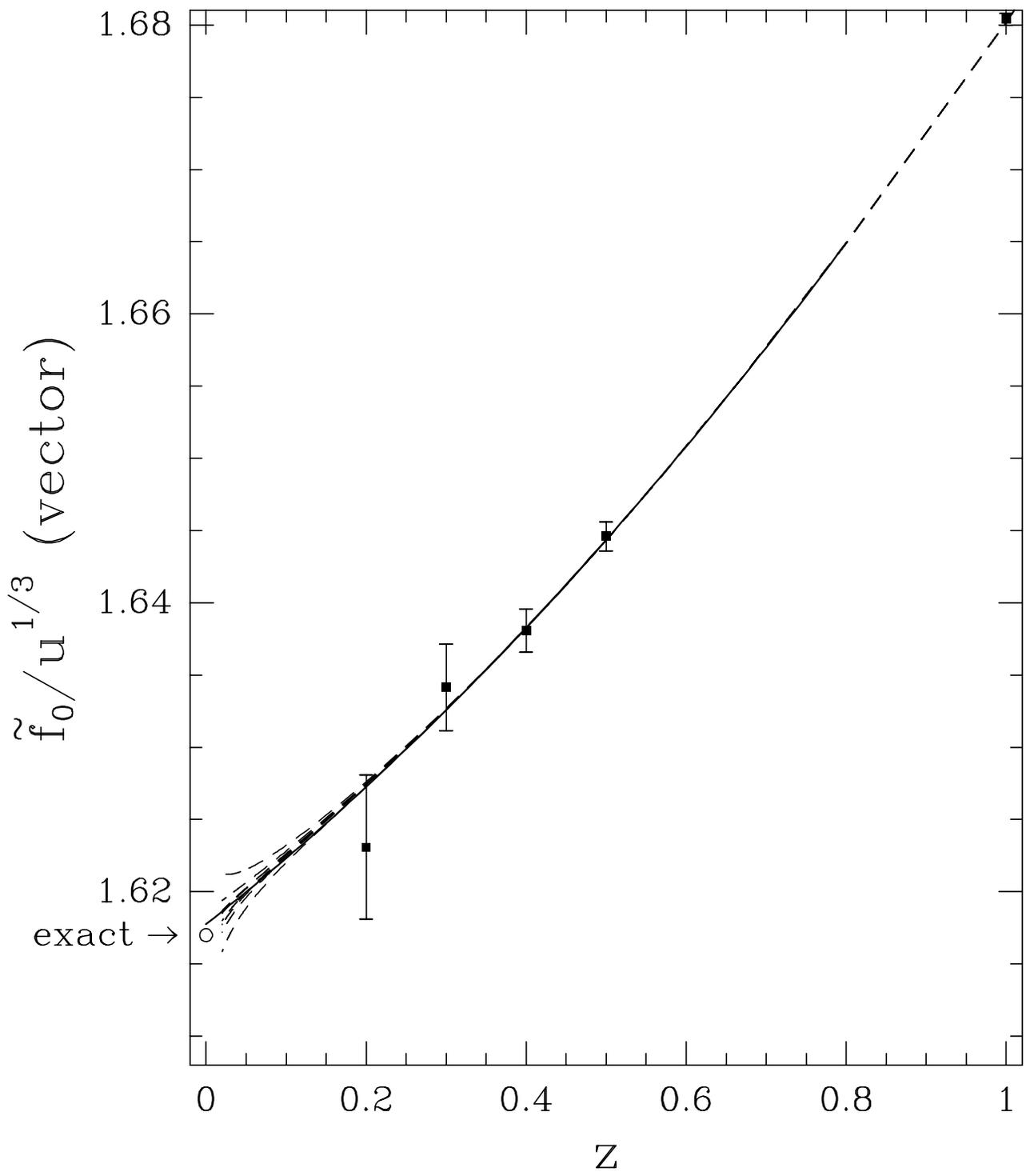
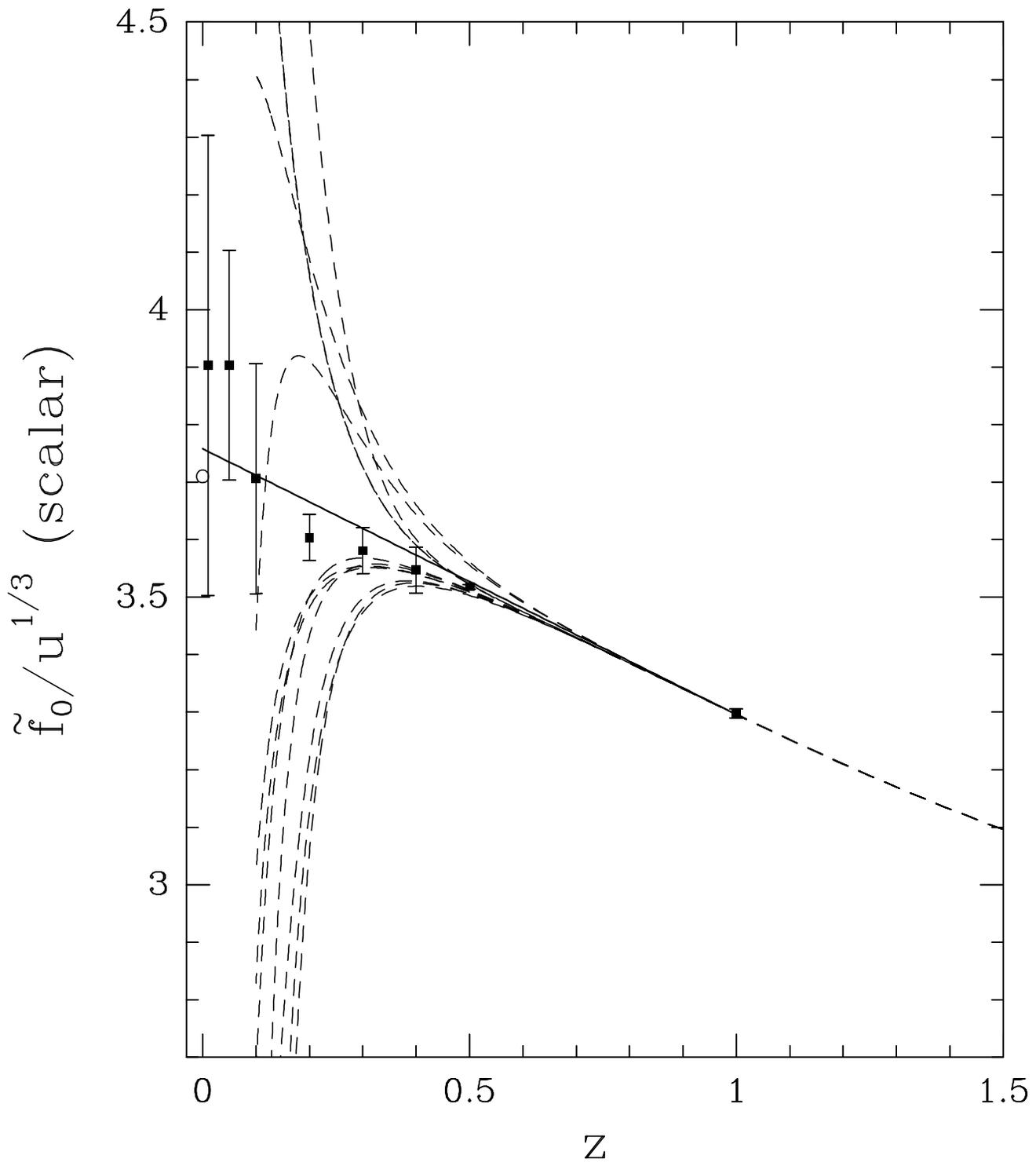
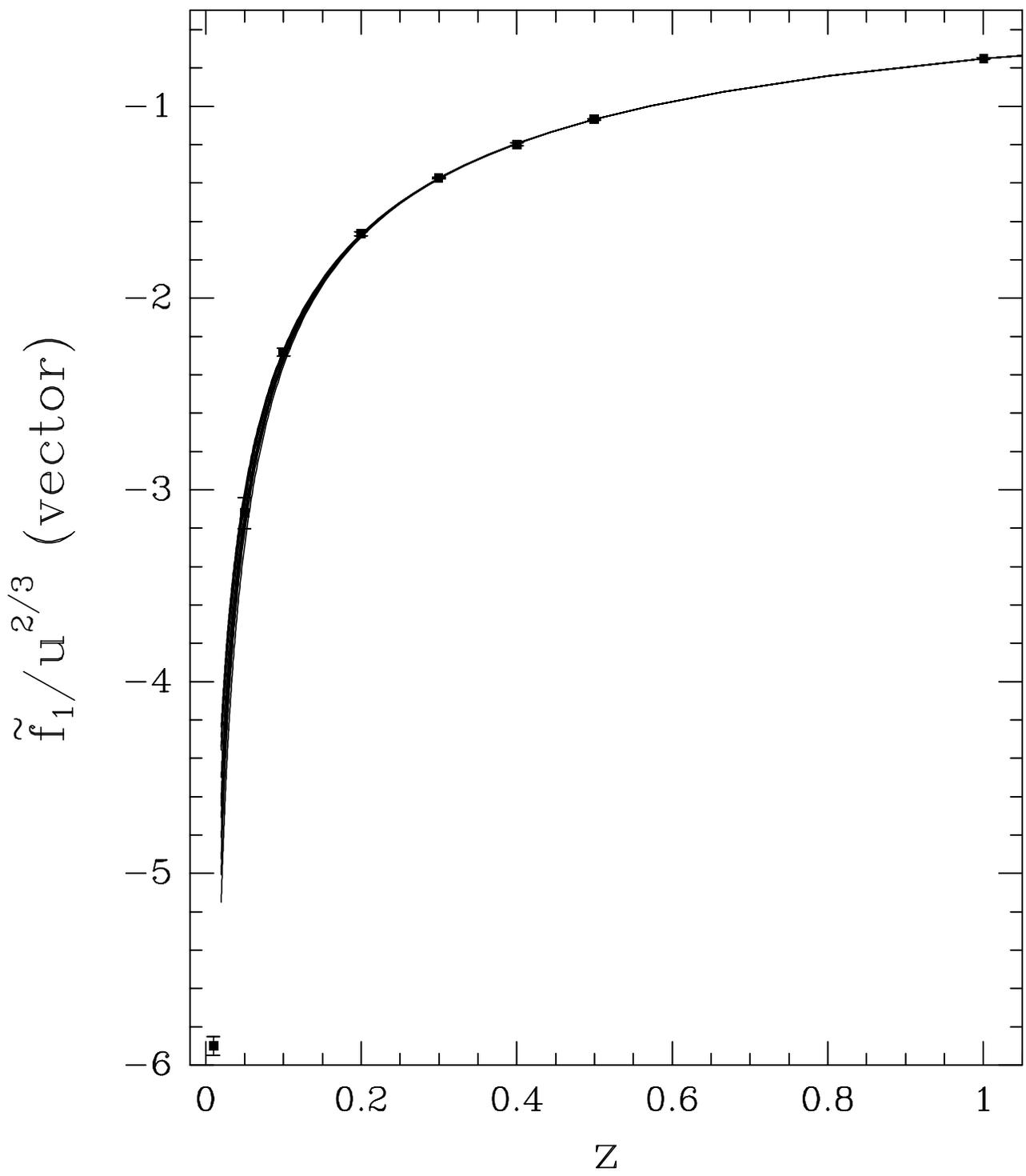
\end{document}